# Defining a classification system for augmentation technology in socio-technical terms

Isabel Pedersen
Faculty of Social Science and
Humanities
Ontario Tech University
Oshawa, Canada
Isabel.pedersen@ontariotechu.ca

Ann Hill Duin
Writing Studies Department
College of Liberal Arts
University of Minnesota
Minneapolis, USA
ahduin@umn.edu

*Abstract*—**This short paper provides a means to classify augmentation technologies to reconceptualize them as socio-technical, discursive and rhetorical phenomena, rather than only through technological classifications. It identifies a set of value systems that constitute augmentation technologies within discourses, namely, the intent to enhance, automate, and build efficiency. This short paper makes a contribution to digital literacy surrounding augmentation technology emergence, as well as the more specific area of AI literacy, which can help identify unintended consequences implied at the design stages of these technologies.**

## I. INTRODUCTION

Augmentation technologies have been emerging over the past 20 years driven by corporate development, university research, military-industrial complex development, increased data availability, new AI techniques, biotech, upgraded computing power and maturing digital architectures. AI has recently accelerated their emergence through machine-learning algorithms, natural language processing, and predictive models that inform the design of technologies. Different sectors and stakeholders use the term Augmentation Technology to explain, justify, and promote the adoption of new technologies; however, as a cohort they do not use the term consistently. We conducted a search of the Association of Computing Machinery (ACM) digital library using the terms *augmentation technologies*, *human augmentation*, and *enhanced humans* and found 276 articles. Neither augmentation technologies nor enhancement technologies are listed in the IEEE Taxonomy: A Subset Hierarchical Display of IEEE Thesaurus Terms nor the 2012 ACM Computing Classification System. We discovered that the articles do not use the term consistently in content or categories. The term augmentation is used ambiguously rather than as a cohesive category describing ways that humans are/will be augmented.

This short paper provides a means to classify augmentation technologies in human-centric terms that reconceptualizes augmentation technologies as socio-technical, discursive and rhetorical phenomena, rather than only through technological classifications. Computer science is oftentimes criticized for presenting technology in objective neutral terms, to the detriment of citizens in the public sphere [1]. Humanities and social science provide methods to presuppose the socio-ethical consequences of technology through critical questioning to understand which stakeholders might be impacted by their deployments. This short paper makes a contribution to digital literacy [2] [3] surrounding augmentation technology emergence, as well as the more specific area of AI literacy to help facilitate AI principles of transparency and explainability in emerging augmentation technologies. It proposes four usable categories as a means to classify augmentation technology in discursive rhetorical terms geared to scientific and technical professionals as well as developers. Moreover, it addresses artificial intelligence and automation as well as ethical and human values in emerging technology.

As background, prior work introduced a framework for scholars and instructors to investigate and plan for social, digital literacy, and civic implications of collaborative, algorithmic, and autonomous writing futures [4]. It emphasizes the importance of cultivating the ability to write, work and interact with augmentation technologies; this includes working with such non-human agents, understanding the impact of algorithms and AI on work and writing, accommodating the unique relationships with autonomous agents, and planning for ongoing disruption. Moreover, it calls for greater depth for understanding augmentation technologies, and amid enhancement of personal and professional capabilities, building ability to articulate its human benefits, risks, and relevance. Such understanding begins with defining augmentation technology in human-centric terms.

This paper also affiliates its goals with the turn toward recognizing Public Interest Technology (PIT). PIT articulates technology design based on citizen needs, "As technology has become an integral part of our everyday life, it has also become intertwined with the public interest. How might citizen rights be protected in the face of emerging technologies? What has to happen? What has to change? How might technology be used to improve civic operations, and at the same time, lessen the controversies of unintended consequences when values like privacy are ignored" [5]. Concerning AI literacy and ethics, similar projects include "A New AI Lexicon" by AI Now Institute that involves "a process of reimagining AI futures by rearticulating the scope and boundaries of critical AI" toward

This project draws on funding from the Canada Research Chairs Program (CRC).

ethical and human-centred ends [6]. Others are writing about augmentation technologies as discursive and rhetorical phenomena. Zizi Papacharissi explores "how human augmentic technologies and artificial or sentient forms of intelligence can be used to enable, reimagine, and reorganize how we understand our selves, how we conceive the meaning of "human", and how we define meaning in our lives" [7]. Heidi A. McKee and James E. Porter concentrate on AI ethics, using the lens of rhetorical context to ground human and nonhuman interaction [8].

## II. METHODOLOGY

We use discourse analysis to classify augmentation technologies in human-centric terms in the social contexts through which they are framed. The word *discourse* comes from the Latin, *discursus,* "exchange of ideas" or discussion. Discursive analysis looks beyond single, isolated texts to define meaning within contexts of human and societal use. We are interested in the way ideas, value systems, and persuasive language (rhetoric) organize beliefs about augmentation from within expert and non-expert spheres. We sampled and analyzed texts from several discourses: (1) academic research articles and conference proceedings in the ACM digital library using search terms such as *augmentation technology*, *human augmentation* and *human enhancement*; (2) texts from corporate websites and corporate news sources promoting, advertising, or celebrating concepts, products, or innovations in augmentation technology; and (3) news from global, private research and advisory firms, such as that of Gartner Group, which tracks and reports on emerging technology through its public news and annual Hype cycle. In this discursive category, we include popular news sources, such as Science which publishes science-related news.

## III. RESULTS

The discourse analysis reveals that human augmentation technologies are delineated according to both technological advancement and a purposeful assumption that humans will be improved through some means. For example, in a special call for papers on robotics for human augmentation from Science Robotics, the author frames the concept: "Fundamental advances in robotics will extend human sensory experience, physicality, and cognition to an unprecedented level. Aided by augmentation technology, the future human will be stronger, faster, less prone to injury, and more productive" [9]. At the same time, augmentation technologies often function through a discourse of futurism, "the future human" [9], that justifies their adoption in abstract terms. One article predicts: "Technology is now on the cusp of moving beyond augmentation that replaces a human capability and into augmentation that creates superhuman capabilities" [10]. One goal of digital literacy is to ground promoted technologies in the early phases of design in order to question the possibility of unintended consequences.

Another finding is that aaugmentation technologies are embodying; they add to the body (or ambient environment around the body). Embodied computing is defined as those technologies that "exist in topographical [on the body], visceral [in the body], and ambient relationships with the body [around the body]" [11]. Defining augmentation technologies involves describing their physical integration with humans in dynamic movement.

We identify a set of three value systems that constitute augmentation technologies within discourses, namely, the intent to enhance, automate, and build efficiency:

- Enhancement – We specifically identify physical, cognitive, sensory, and emotional enhancement as a range of subcategories (see Table 1). These subcategories help us classify nearly all types of augmentation. More importantly, these provide a working terminology for technical communicators that allows for discussion of the intent to augment a process without being limited to discussion of a specific piece of hardware. For example, cognitive enhancement is realized through brain implants, wearable skin patches, and headsets; however, the concept itself, the intent to enhance might be more important than its material instantiation.

- Automation – Augmentation technology assumes a degree of automation as a value, a logic imposed upon it largely by the mass adoption of Artificial Intelligence (AI). This assumption operates as an assumed or invisible rhetoric used to drive the development and adoption of these technologies, which is rarely questioned.

- Efficiency – Humans, human work, and human experiences are reconceptualized in terms of usefulness or avoiding wasting energy, time, or resources. Here we note a shift in this paradigm. While physical enhancement might have always assumed efficiency as a goal, enhancement in thinking, sensing, and feeling is new territory in discursive deployments, and it may lead to harms that have gone uninvestigated.

The burgeoning augmentation technologies market masks complex sociotechnical tradeoffs that emerge from the intent to enhance, automate, and build efficiency. Scientific and technical professionals must understand these value systems as part of their work to develop usable and ethical content. For example, a business will frame an augmentation technology as a business solution without conveying critical information about the concessions that users must make in terms of giving up their data in exchange for access and use.

Table 1 (at the end of the paper) provides an overview of the four enhancement subcategories that we use to classify augmentation technology. For each category we provide a set of synonyms used across different sectors and stakeholders when defining the augmentation as well as the general goals for its use. For goals, justifications, or rhetorical rationale, we summarize the intent to design a technology. We provide an example selection of augmentation technologies including emerging hardware, software, and applications.

## IV. ENHANCEMENT SUBCATEGORIES

### A. Cognitive enhancement

Goals for cognitive enhancement include the perceived need to be smarter, more knowledgeable, think faster, remember more, know more, learn faster, learn more efficiently, edit dysfunctional memory, or be more reasonable.

These terms are always changing, but they cover concepts related to human cognitive abilities. Regarding thinking enhancement, in the 1960s Doug Engelbart explored how cognitive abilities could be enhanced through rapid access to information and increasing user comprehension, developing an early intelligent user interface for efficiency called the Intelligence Amplifying System of Tools [12]. Many researchers have since proposed and developed various interfaces for cognitive and memory enhancement [13] [14] [15] [16]. A recent paper proposes the "Wearable Reasoner: Towards Enhanced Human Rationality Through A Wearable Device With An Explainable AI Assistant" [16]. It also bases its justification on efficiency and automation stating the need "due to the fact that the human mind is rather limited in time, knowledge, attention, and cognitive resources" [16].

Business websites, indicating a discourse geared to non-experts, define and classify cognitive enhancement in terms of human and non-human collaboration. Gartner Inc. defines "Augmented intelligence" as "a design pattern for a human-centered partnership model of people and artificial intelligence (AI) working together to enhance cognitive performance, including learning, decision making and new experiences" [17]. In this way, cognitive enhancement comes about as a result of partnership in which people and AI work together through automation of processes. Another example defines augmented intelligence as "an alternative conceptualization of artificial intelligence," also focusing on AI's assistive role, and "emphasizing the fact that cognitive technology is designed to enhance human intelligence rather than replace it" [18].

## B. *Sensory Enhancement*

Goals for sensory enhancement include the need to experience more through augmentation of the senses or perception, to augment or reconstruct one's reality through digital representations, to escape one's sensory reality, and to focus better.

We use the term immersive technologies to mean forms of sensory augmentation that are often visual but can also include auditory or tactile immersion. They include virtual reality, augmented reality, mixed reality and immersive videos. Traditionally, Virtual Reality VR immerses users in a completely computer-generated 3D environment, while Augmented AR combines virtual components as an overlay over the real world, a sensory augmentation.

Justifications for sensory augmentation are myriad. Inventor Arnav Kapur says that using his AlterEgo device will help people to become "better at being human," [19] cloaking it with the augmentation rhetoric that defines the field. AlterEgo is a wearable AI device "with the potential to let you silently talk to and get information from a computer system, like a voice inside your head" [19]. It enhances speaking by transcribing "words that the user verbalizes internally but does not actually speak aloud" [20]. It is a "silent speech" augmentation created by sensing the neuromuscular signals of the jaw, the speaker's bone conduction that can only be interpreted by the wearer.

## C. *Emotional Enhancement*

Goals for emotional enhancement of humans include the desire to feel that devices understand one's emotions, to control one's emotion, or even be happier or more fulfilled.

Countless domains now work toward ambitious goals to solve world problems upon the expectation that AI will someday achieve human-like empathy to enhance humans. Critical to emotional enhancement is that ethical design practices must underpin all work in this area to ensure that human rights are always respected. Researchers in AI ethics, robot ethics, and philosophy are delineating value-based frameworks for embedding synthetic emotions in augmentation technologies. The IEEE Global Initiative on Ethics of Autonomous and Intelligent Systems includes a chapter on Affective Computing in the first edition of *Ethically-Aligned Design* [21]. Kate Crawford writes, "For the world's militaries, corporations, intelligence agencies, and police forces, the idea of automated affect recognition is as compelling as it is lucrative. It holds the promise of reliably filtering friend from foe, distinguishing lies from truths, and using the instruments of science to see into interior worlds" [1]. However, detecting and analyzing facial expressions -- automated affect detection systems -- are being deployed in industry despite "a lack of substantial scientific evidence that they work" [1]. She writes "Instead of trying to build more systems that can group expressions into machine-readable categories, we should question the origins of those categories themselves, as well as their social and political consequences" [1].

## D. *Physical Enhancement*

Goals for physical enhancement include the opportunity to be stronger, faster, live longer, live healthier, work longer, physically work more efficiently, and play sports better.

One augmentation technology, an exoskeleton, is defined as "a body-worn human enhancement technology that takes the form of a suit" [22]. Exoskeletons, sometimes called 'wearable robotics,' are slated to dramatically enhance human capabilities proposed to combine with AI, Internet of Things, AR, and other technologies. Another example is digital fiber designed at MIT that senses, stores, analyzes, and infers activity after being sewn into a shirt. One team describes this work as "the first realization of a fabric with the ability to store and process data digitally, adding a new information content dimension to textile and allowing fabrics to be programmed literally" [23]. Representing on-body AI, the fiber is able "to determine with 96% accuracy what activity the person wearing it was engaged in… With this analytic power, the fibers someday could sense and alert people in real-time to health changes like a respiratory decline or an irregular heartbeat, or deliver muscle activation or heart rate data to athletes during training" [23].

## V.   CONCLUSION

We have classified augmentation technologies as socio-technical, discursive and rhetorical phenomena. Augmentation technology involves the intent to enhance, automate, and build efficiency. The scope for human augmentation technologies encompasses the goal to enhance cognitive, sensory, emotional and physical capabilities of humans. We follow technology emergence through phases of design, adoption and adaptation by monitoring discursive ordering and activity. A discourse of futurism, how humans will be enhanced in future, functions as a persuasive instigator across the language.

The next step is to reveal (and question) unintended consequences implied at the design stages, an important function of digital literacy and AI literacy. As part of this presentation, we will provide an overview of the larger Humane Futures project and associated framework as a guide for reframing scholarship and practice to promote digital and AI literacy surrounding the ethical design, adoption, and adaptation of augmentation technologies. This project includes a digital research repository that will house our findings, using the classification system as the basis for metadata.

TABLE I.    TABLE 1: SCOPE OF AUGMENTATION TECHNOLOGIES

| | Cognitive enhancement | Sensory enhancement | Emotional enhancement | Physical enhancement |
|---|---|---|---|---|
| Synonyms | Intelligence amplification, memory amplification, machine augmented intelligence | Perceptual enhancement, Immersive enhancement | Affective enhancement, Emotion manipulation, Emotion regulation | Augmentation technology |
| Goals, justifications, or rhetorical rationale | Be smarter, more knowledgeable, think faster, remember more, know more, learn faster, learn better (efficiently), edit memory, be more reasonable | Experience more, experience more through the manipulation of the senses, augment one's senses to construct an enhanced reality, escape one's sensory reality, focus better | Feel more and different emotions or feel less emotion, control emotion, be happier | Be stronger, faster, live longer, live healthier, work harder, work longer, physically work more efficiently, play sports better |
| Selection of augmentation technologies hardware, software | Brain-computer interaction, Human-AI interaction/ collaboration, AI Virtual assistants | Augmented reality, virtual reality, 3D immersive space, Cochlear implants, (Auditory) smart translation, tactile displays | Affective computing, Emotion detection, Emotion decoding technology, Biometrics | Exoskeletons/exosuits, Smart shirts, Programmable fabric, Wearable robotics, Smart prosthetics |